\documentclass[final,3p,times,onecolumn,sort&compress,hidelinks]{elsarticle}

\usepackage{hyperref}
\usepackage{bm, color}
\usepackage{graphicx}
\usepackage[utf8]{inputenc}
\usepackage{epstopdf}

\journal{Physics Letters B}

\begin{document}

\begin{frontmatter}

\title{Testing the universality of the Collins function in pion-jet production at RHIC}


\author[inst1,inst2]{Umberto D'Alesio\corref{cor1}}\ead{umberto.dalesio@ca.infn.it}
\author[inst2]{Francesco Murgia}\ead{francesco.murgia@ca.infn.it}
\author[inst3]{Cristian Pisano}\ead{cristian.pisano@pv.infn.it}
\address[inst1]{Dipartimento di Fisica, Universit\`a di Cagliari, Cittadella Universitaria,
 I-09042 Monserrato (CA), Italy}
\address[inst2]{INFN, Sezione di Cagliari, Cittadella Universitaria, I-09042 Monserrato (CA), Italy}
\address[inst3]{Dipartimento di Fisica, Universit\`a di Pavia, \& INFN, Sezione di Pavia, Via A.~Bassi 6, I-27100, Pavia, Italy}
\cortext[cor1]{Corresponding author}

\begin{abstract}
By adopting a generalised parton model approach at leading order in QCD, including spin and intrinsic parton motion effects, we study the Collins azimuthal asymmetries for pions within a large-$p_T$ jet produced at mid-rapidity in polarised hadronic collisions. Using available information on the quark transversity distributions and the pion Collins functions, as extracted from semi-inclusive deeply inelastic scattering and $e^+e^-\to h_1 h_2\, X$ processes, we compute estimates for the Collins asymmetries in kinematical configurations presently investigated at RHIC by the STAR Collaboration. Collins-like asymmetries, involving linearly polarised gluons, are also considered.
Our predictions, compared against available preliminary data, show a very good agreement, even if some discrepancies, to be further scrutinized both theoretically and experimentally, appear in the transverse momentum dependence of the Collins asymmetry. These results are in favour of the predicted universality of the Collins function and of a mild, if any, evolution with the hard scale of the asymmetries.
\end{abstract}

\begin{keyword}
QCD phenomenology\sep Polarization effects\sep Transverse momentum dependent distributions \sep Universality
\end{keyword}

\end{frontmatter}


\section{Introduction}
\label{intro}

Transverse single-spin asymmetries (SSAs) are a long-standing challenge and a striking issue for collinear leading-twist perturbative QCD. Well-known examples are: the large transverse polarisation of $\Lambda$ hyperons produced in unpolarised
$pp$, $pA$ collisions; the sizable pion SSAs measured in polarised $pp$ collisions, first at fixed target
experiments and recently at RHIC, at much larger c.m.~energies and pion transverse momentum;
the azimuthal asymmetries measured in semi-inclusive deeply inelastic scattering (SIDIS) and in $e^+e^-$
collisions. These observables and their understanding are indeed crucial for a full knowledge of
the nucleon structure, concerning in particular parton orbital motion and angular momentum, and the consequences for spin sum rules and the violation of helicity selection rules in high-energy exclusive processes.
It is by now clear that the spin of the proton cannot be explained simply by the sum of the spins of its
constituents, including sea quarks and gluons, and that a crucial role could be played by parton orbital angular momentum.
The so-called transverse-momentum-dependent~(TMD) approach, including spin and intrinsic parton motion, is by now accredited as one of the theoretical formalisms able to account for many of these spin effects, at least in certain kinematical configurations (see e.g.~Refs.~\cite{D'Alesio:2007jt,Barone:2010zz,Aschenauer:2015ndk} for general reviews on SSAs and TMDs and Refs.~\cite{Collins:2011zzd,GarciaEchevarria:2011rb,Collins:2017oxh} for the latest theoretical developments).

For an alternative approach, extending the usual collinear scheme with inclusion of higher-twist effects and gluon-quark-gluon correlators, see for example Refs.~\cite{Kouvaris:2006zy,Kanazawa:2015ajw} and references therein.

In the TMD approach the spin asymmetries are ultimately due to TMD parton distribution (TMD-PDF) and fragmentation (TMD-FF) functions. The most relevant ones are the Sivers distribution function~\cite{Sivers:1989cc} and the Collins fragmentation function~\cite{Collins:1992kk}.
The leading-twist TMD Collins FF describes the asymmetry in the azimuthal distribution of an unpolarised hadron around the direction of motion of the transversely polarised fragmenting parent quark.
It is chiral odd and naively T-odd, and is nonvanishing only if the transverse motion of the produced hadron (w.r.t.~the quark direction of motion) is explicitly taken into account.

According to TMD factorisation theorems, valid for two energy-scale processes like SIDIS, $e^+e^-$ annihilation and Drell-Yan processes, the Collins FF is expected to be universal and process independent~\cite{Collins:2004nx}. It is responsible for the azimuthal correlation in the distribution of two hadrons produced in opposite jets in $e^+e^-$ collisions and for a specific azimuthal modulation measured in SIDIS with a transversely polarised target.
In inclusive single-hadron production in polarised $pp$ collisions the situation is definitely more involved,
both theoretically and phenomenologically, since, within a generalised parton model (GPM) approach (a parton model with inclusion of spin and transverse momentum effects), the Sivers and Collins effects cannot be disentangled (see Refs.~\cite{Anselmino:2012rq,Anselmino:2013rya}). Moreover, a proof of factorisation in terms of TMDs for such processes (where only one energy scale is present) is still lacking~\cite{Rogers:2013zha} and potential factorisation breaking effects could be expected~\cite{Collins:2007nk}. Testing of universality is therefore a very non-trivial topic, sensitive to important aspects of QCD.

On the other hand, the study of the azimuthal distribution of leading hadrons produced in the fragmentation of a large transverse momentum jet offers a unique opportunity for studying TMDs as well as factorisation breaking effects in hadronic collisions and, by comparison with SIDIS and $e^+e^-$ annihilations, get information on the process dependence of the TMD functions. For instance, a detailed study of the Sivers function in the same process has been presented in Ref.~\cite{D'Alesio:2011mc}. In this respect these observables represent a very powerful testing ground of our understanding of SSAs and TMD effects.

The Collins asymmetry for the $p^\uparrow p\to {\rm jet}\;\pi\,X$ process at large rapidity was first considered in
Ref.~\cite{Yuan:2007nd}, accounting for parton transverse momenta only in the fragmentation process and giving at the same time the proof of factorisation (at least at the lowest order).
The approach was generalised in Ref.~\cite{D'Alesio:2010am} including parton motion also in the initial distributions, within a TMD phenomenological model, and providing the full leading-twist structure for azimuthal asymmetries. A detailed analysis for several interesting asymmetries in kinematical configurations reachable at RHIC was also presented, while subsequently, in Ref.~\cite{DAlesio:2013cfy}, a short dedicated review and further phenomenological studies were collected. Notice that, while the complete $k_\perp$ treatment adopted in the GPM approach assumes factorisation and remains to be checked experimentally, it allows a much richer phenomenological analysis, also in view of testing potential factorisation breaking effects. It is nevertheless true that for the Collins asymmetry this model would give results similar to those obtained adopting the collinear approach for the jet production mechanism of Ref.~\cite{Yuan:2007nd}.

Since then, the first preliminary data for the Collins asymmetry in the mid-rapidity region have been released by the STAR Collaboration, both at $\sqrt{s}=$ 200 and 500 GeV~\cite{Adkins:2016uxv}.

Together with data from SIDIS and $e^+e^-$ collisions, these results allow for the first time a direct test of the universality of the Collins function (as well as of transversity) and the effective role of its scale dependence.

Therefore, in this letter we apply the TMD GPM approach to the Collins asymmetry in pion-jet production (as presented in Ref.~\cite{D'Alesio:2010am}), \emph{keeping fixed} the parameterizations of the transversity and Collins functions as obtained by fitting the SIDIS and $e^+e^-$ results. By imposing the same kinematical cuts as adopted at RHIC, we give predictions for the Collins asymmetry in $p^\uparrow p\to {\rm jet}\;\pi\,X$ processes and compare our results with the new preliminary STAR data, looking for information on universality and scale dependence of the Collins function. For completeness, we will also consider the gluon Collins-like asymmetry, due to the convolution of the TMD distribution of linearly polarised gluons inside a transversely polarised proton with the TMD fragmentation function of linearly polarised gluons into an unpolarised hadron. In such a case we will show how present preliminary data~\cite{Drachenberg:2014txa,Drachenberg:2014jla} could help in constraining the \emph{product} of these two completely unknown TMDs. A separate and more direct extraction of the linearly polarized gluon distribution would be possible by looking at dijet and heavy-quark pair production in electron-proton collisions at a future Electron Ion Collider~\cite{Boer:2016fqd}.

The paper is organized as follows: in section~\ref{formalism} we recall the basic ideas of the formalism, skipping all details of the calculations that can be found in Ref.~\cite{D'Alesio:2010am}; in section~\ref{results} we present our theoretical estimates and compare them with the available preliminary experimental results; conclusions and open issues are gathered in section~\ref{conclusions}.

\section{Formalism}
\label{formalism}

The azimuthal moment of the Collins asymmetry for the $p^\uparrow p\to {\rm jet}\,\pi\, X$ process is defined as follows:
\begin{equation}
A_N^{\sin(\phi_S-\phi^H_\pi)}(\bm{p}_{\rm j},z,k_{\perp \pi}) =
2\,\frac{\int\,d\phi_S \,d\phi_{\pi}^H\,\sin(\phi_S-\phi_\pi^H)\,[\,d\sigma(\phi_s,\phi_\pi^H)-
d\sigma(\phi_s+\pi,\phi_\pi^H)\,]}
{\int\,d\phi_S\, d\phi_{\pi}^H\,[\,d\sigma(\phi_s,\phi_\pi^H)+d\sigma(\phi_s+\pi,\phi_\pi^H)\,]}\,,
\label{eq:an-col}
\end{equation}
where $d\sigma(\phi_S,\phi_\pi^H)$ is a shorthand notation for the invariant differential cross section
\begin{equation}
\frac{E_{\rm j}\,d\sigma^{p(S,\phi_S) p\to {\rm jet}\,\pi(\phi_\pi^H)\,X}}
{d^3\bm{p}_{\rm j}\,dz\, k_{\perp\pi}\,dk_{\perp\pi}\,d\phi_\pi^H\,d\phi_S}\,,
\label{eq:dsig}
\end{equation}
and the numerator in Eq.~(\ref{eq:an-col}), taking into account that another allowed term becomes negligible upon integration over the partonic azimuthal phases, is schematically given as
\begin{equation}
N[A_N^{\sin(\phi_S-\phi_\pi^H)}] \sim h_1^q(x_a,\bm{k}_{\perp a}^2) \otimes
 f_1(x_b,\bm{k}_{\perp b}^2)\otimes \Delta\hat\sigma \otimes H_1^{\perp\, q}(z,\bm{k}_{\perp \pi}^2)\,.
\label{eq:Coll2}
\end{equation}
Notice that in the denominator of the SSA, that is twice the unpolarised cross section, we include all kind of partons, quarks and gluons.

{}For all details we refer to Ref.~\cite{D'Alesio:2010am}. Here we only recall that $\bm{p}_{\rm j}$ is the jet three-momentum, that will be expressed in terms of its pseudorapidity $\eta_{\rm j}$
and transverse momentum $p_{{\rm j}T}$; $z$ is the fraction of the jet momentum carried by the observed pion; $\bm{k}_{\perp\pi}$ is the transverse momentum of the pion with respect to the parent parton
(the jet in our leading-order approach), and $\phi_\pi^H$ is the azimuthal angle of the pion momentum, measured in the jet helicity frame; $S$ is the (transverse) polarisation vector of the initial proton beam, forming an angle $\phi_S$ with the jet production plane (the $x$-$z$ plane) in the $pp$ c.m.~reference frame.
The azimuthal factor $\sin(\phi_S-\phi^H_\pi)$ in the numerator of Eq.~(\ref{eq:an-col}) singles out the Collins contribution, given in Eq.~(\ref{eq:Coll2}) as the convolution of the TMD transversity distribution for quarks inside the polarised proton, $h_{1}^q$, with the TMD unpolarised distribution for the partons (quarks and gluons) inside the unpolarised proton, $f_1$, the Collins function in the fragmentation process, $H_1^{\perp q}$, and the partonic spin transfer, $\Delta\hat\sigma$.

If, alternatively, one uses as azimuthal weight the factor $\sin(\phi_S-2\phi^H_\pi)$ then the so-called Collins-like contribution is singled out, as discussed briefly in the sequel. This term is related to the convolution of the TMD distribution for linearly polarised gluons in the polarised proton (which vanishes in the collinear configuration) with the unpolarised TMD distribution in the unpolarised proton and the Collins-like gluon FF in the jet fragmentation process (that parameterises the fragmentation of linearly polarised gluons into an unpolarised hadron).
Other possible azimuthal asymmetries, like the Sivers effect, will not be considered in this paper~\cite{D'Alesio:2010am}.

In the calculations, the TMD PDFs and FFs are parameterised using a simple form, where the light-cone momentum fraction and transverse momentum dependences are factorised, and only the DGLAP QCD evolution of the collinear parts, choosing $p_{{\rm j}T}$ as factorisation scale, is taken into account. The main reason is that a proper TMD evolution framework for such a process is still not available and could be potentially different from what is usually adopted in processes like SIDIS. Moreover, it is interesting by itself to see to what extent the ordinary collinear evolution is able to describe the available data, which, even if taken at very different $p_{{\rm j}T}$ values, do not show any significant scale dependence~\cite{Adkins:2016uxv}. Further study in this direction would be certainly helpful.

The transverse-momentum dependent part is taken to be Gaussian (times proper $k_\perp$ factors) and flavor independent.
All free parameters entering the parametrisation of the transversity distribution and of the Collins functions are
fixed by fitting experimental data for the Collins asymmetry in SIDIS pion production by the HERMES and Compass Collaborations, and for two-hadron azimuthal correlations in $e^+e^-\to \pi\pi\,X$ processes by the Belle and BaBar Collaborations.

No additional free parameter is introduced in our analysis, and the theoretical curves, compared with the preliminary STAR experimental data on the Collins asymmetry in $p^\uparrow p\to {\rm jet}\,\pi\,X$, are to be considered as direct predictions based on a TMD factorisation scheme.

In the following we will adopt different sets of the quark transversity and Collins functions,
which we will refer to as the SIDIS~1~\cite{Anselmino:2007fs}, the SIDIS~2~\cite{Anselmino:2008jk} sets and the 2013 fit~\cite{anselmino:2013vqa}. The reason of this choice is that the SIDIS~1 and SIDIS~2 sets are those used in our first study~\cite{D'Alesio:2010am}: in this respect, by keeping them in the present analysis we emphasise its aspect of being a natural extension of our former work, with predictions somehow obtained directly from there. While they are well representative of the available extractions of these TMDs and their uncertainties, for the sake of completeness and to keep our study up-to-date we have considered the more recent 2013 fit as well. In such a case in showing our estimates we will also provide the corresponding statistical uncertainty bands, calculated following the procedure described in Appendix A of Ref.~\cite{Anselmino:2008sga}.

A comment on the use, and impact, of the most recent extraction~\cite{Anselmino:2015sxa} of the transversity distribution and the Collins FF will be given below.

These sets, besides some differences in the initial assumptions and in the data used for their extraction, differ in the choice of the collinear fragmentation functions. More precisely, for the SIDIS~1 fit the Kretzer FF set~\cite{Kretzer:2000yf} was adopted, while for the SIDIS~2 and the 2013 fits the DSS FF set~\cite{deFlorian:2007aj} was employed. The latter, in particular, incorporates a more sizeable gluon FF w.r.t.~the corresponding one in the Kretzer set, that could play a role in the process under consideration.

Notice that in all cases the extraction of the quark transversity distribution is constrained only up to $x\simeq 0.3$.

\section{Results}
\label{results}
The STAR Collaboration at RHIC has collected preliminary results on the Collins and Collins-like asymmetries for the
$p^\uparrow p\to {\rm jet}\,\pi\,X$ process in the midrapidity range ($|\eta_{\rm j}| \leq 1$) at c.m.~energies $\sqrt{s}=$ 200 and 500 GeV~\cite{Adkins:2016uxv,Drachenberg:2014txa,Drachenberg:2014jla}.
Table~1 summarises the main kinematical cuts relevant for our phenomenological study.  We have implemented all of them in our calculations. Notice that in our leading-order TMD approach the jet is identified with the final fragmenting parton coming from the hard subprocesses and that the STAR Collaboration adopts the anti-$k_t$ jet reconstruction algorithm. For detailed studies of the transverse momentum distribution of hadrons within a jet see Refs.~\cite{Kaufmann:2015hma,Kang:2017glf}.
We also remark that some of the above cuts induce a $\cos\phi^H_\pi$ azimuthal dependence in the (un)polarised cross sections. We have checked that this dependence cancels out in the asymmetries.

\begin{figure}[t!]
\begin{center}
\includegraphics[width=8.25truecm,angle=0]{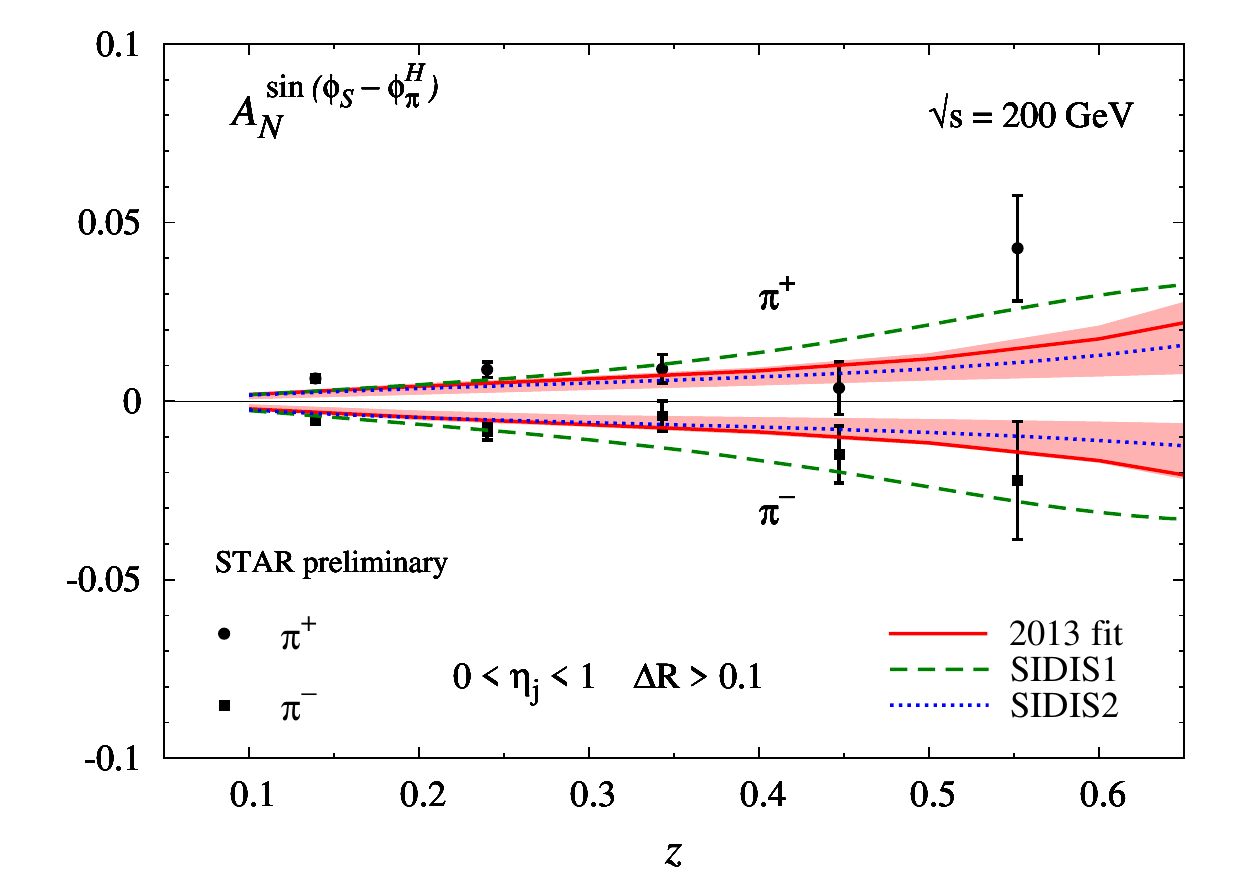}\hspace*{-0.5 cm}
\includegraphics[width=7.5truecm,angle=0]{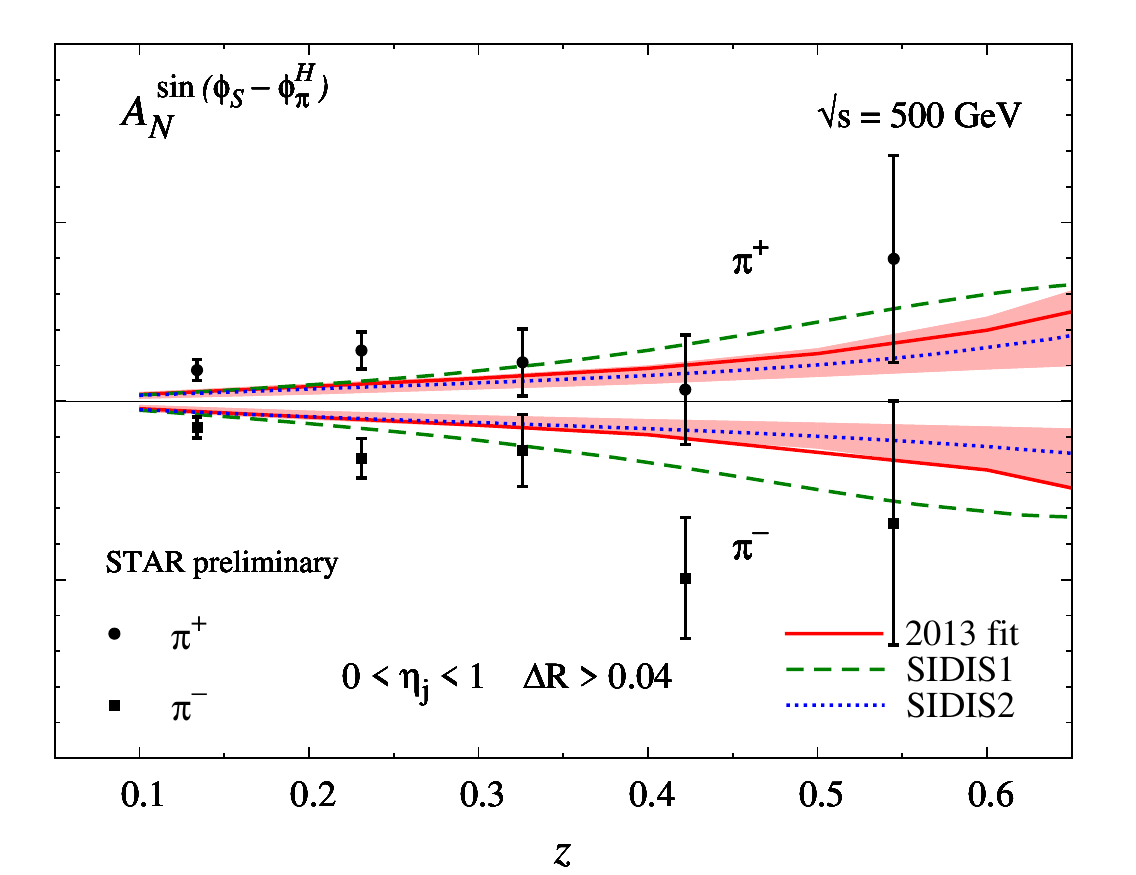}
\caption{The Collins asymmetry $A_N^{\sin(\phi_{S}-\phi_\pi^H)}$ for the process $p^\uparrow \, p\to {\rm jet}\,
\pi^{\pm} \, X$, as a function of $z$, at c.m.~energy $\sqrt{s}= 200$ GeV (left panel) and 500 GeV (right panel), compared with STAR preliminary data~\cite{Adkins:2016uxv}. Estimates are obtained by integrating over the other variables, imposing the experimental cuts as summarised in the legend (see also Table~1) and adopting three different parameterisations: the 2013 fit (red solid lines), with their statistical uncertainty bands, the SIDIS~1 (green dashed lines) and SIDIS~2 (blue dotted lines) parameterisations.}
\label{fig:asy-coll-par-z}
\end{center}
\end{figure}

In Fig.~\ref{fig:asy-coll-par-z} we present our theoretical estimates of the Collins asymmetry for $\pi^\pm$ production at $\sqrt s = 200$~GeV (left panel) and 500~GeV (right panel) as a function of $z$, compared with STAR preliminary data~\cite{Adkins:2016uxv}, integrated over the other kinematical variables and imposing the experimental cuts of Table~1. Our predictions are given for the sets of the transversity distribution and the Collins FF discussed above; in particular, for the 2013 fit we also show their statistical uncertainty bands. The agreement between theory and data is definitely very good.

\begin{table*}[t!]
\centering
\begin{tabular}{|c|c|c|}
\hline
\rule{0pt}{1.2em}
 Kinematical cut    & $\sqrt{s}=200$ GeV & $\sqrt{s}=500$ GeV  \\
\hline
\hline
\rule{0pt}{1.2em}
Jet cone radius & $R=$ 0.6 & $R=$ 0.5 \\
Min$(p_{{\rm j}T})$ & 10.0 GeV & 22.7 (6.0) GeV \\
Max$(p_{{\rm j}T})$ & 31.6 GeV & 55.0 (13.8) GeV \\
Jet pseudorapidity & $|\eta_{\rm j}| \leq 1$ & $|\eta_{\rm j}| \leq 1$ \\
Pion $p_T$ minimum & 0.2 GeV & 0.2 GeV \\
Pion $p_T$ maximum & 30.0 GeV & 30.0 GeV \\
Pion $\Delta R = \sqrt{(\phi_\pi-\phi_{\rm j})^2+(\eta_\pi-\eta_{\rm j})^2}$ & $\Delta R > 0.1$ & $\Delta R > 0.04$ \\
Pion $z=|\bm{p}_\pi|/|\bm{p}_{\rm j}|$ & $0.1 < z < 0.6$ & $0.1 < z < 0.8$ \\
Pion $j_T \equiv k_{\perp\pi}$ & $0.125 < k_{\perp\pi} < 4.5$ GeV &  $0.1 < k_{\perp\pi} < 2.0$ GeV \\
\hline
\end{tabular}
\caption{Kinematical cuts adopted by the STAR Collaboration and used in the present theoretical estimates. The values in round brackets for $\sqrt s = 500$ GeV refer to the cuts for the Collins-like asymmetries.}
\label{cuts}
\end{table*}

Notice that, both at 200 and 500~GeV energies, the average momentum fraction of the quarks inside the polarised proton (as well as inside the unpolarised one) is around 0.2. That means we are probing the valence region of the transversity distribution and, more importantly, the region where it is well constrained by SIDIS data. Almost no energy dependence appears (see also the comments below on the explored kinematical region).
The differences between the estimates obtained adopting the SIDIS~1 set w.r.t.~those obtained with the SIDIS~2 set and the 2013 fit are almost completely due to the fact that the unpolarised cross section computed with the DSS FF set (associated to the SIDIS~2 fit and the 2013 fit, and entering the denominator of the SSA) is bigger than the one computed with the Kretzer set, in the medium and large $z$ region. This, in turn, can be traced back to the large leading-order gluon FF of this set, essentially driven by its extraction from a global fit including also $pp\to \pi X$ data. We have to note that the corresponding next-to-leading-order extraction, not used here, would be more stable.

\begin{figure}[!t]
\begin{center}
\includegraphics[width=8.25truecm,angle=0]{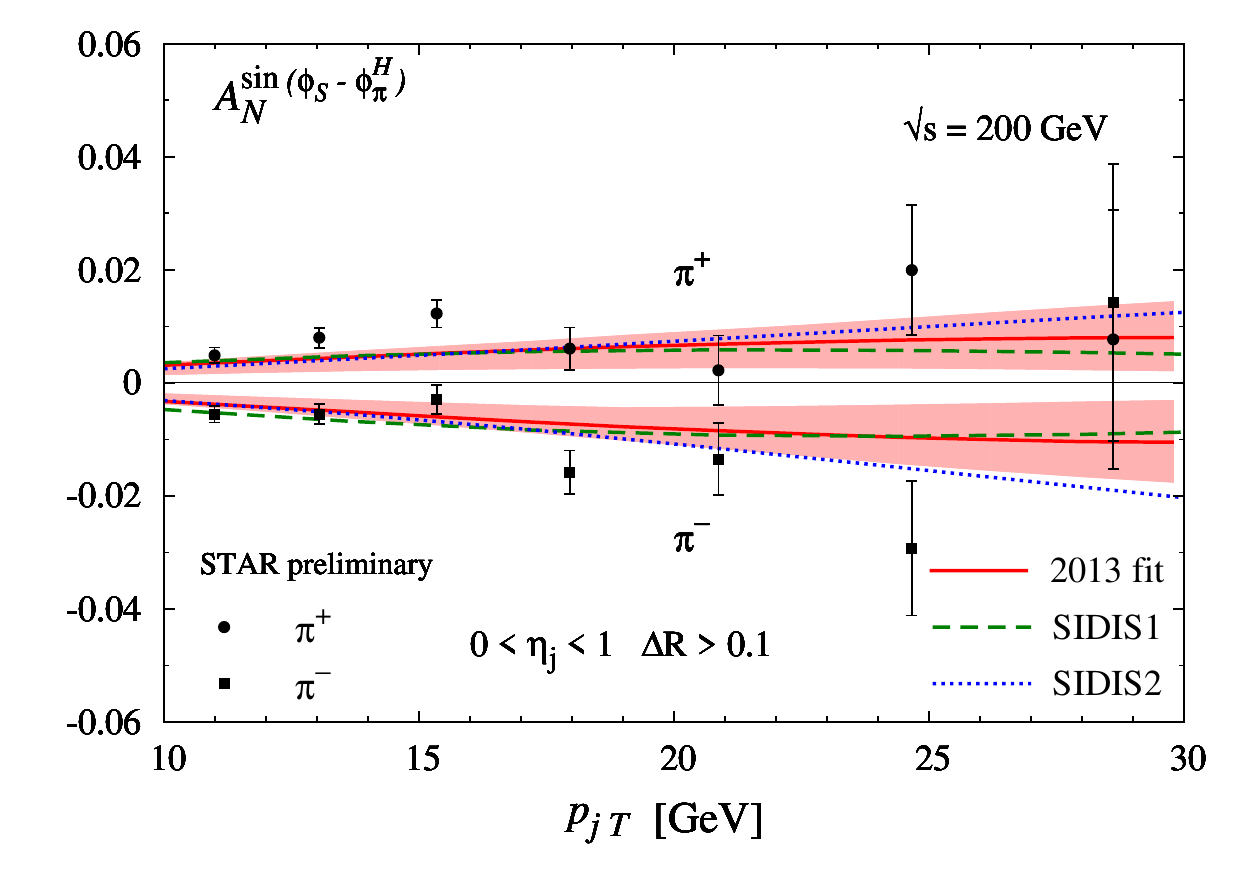}\hspace*{-0.5 cm}
\includegraphics[width=7.5truecm,angle=0]{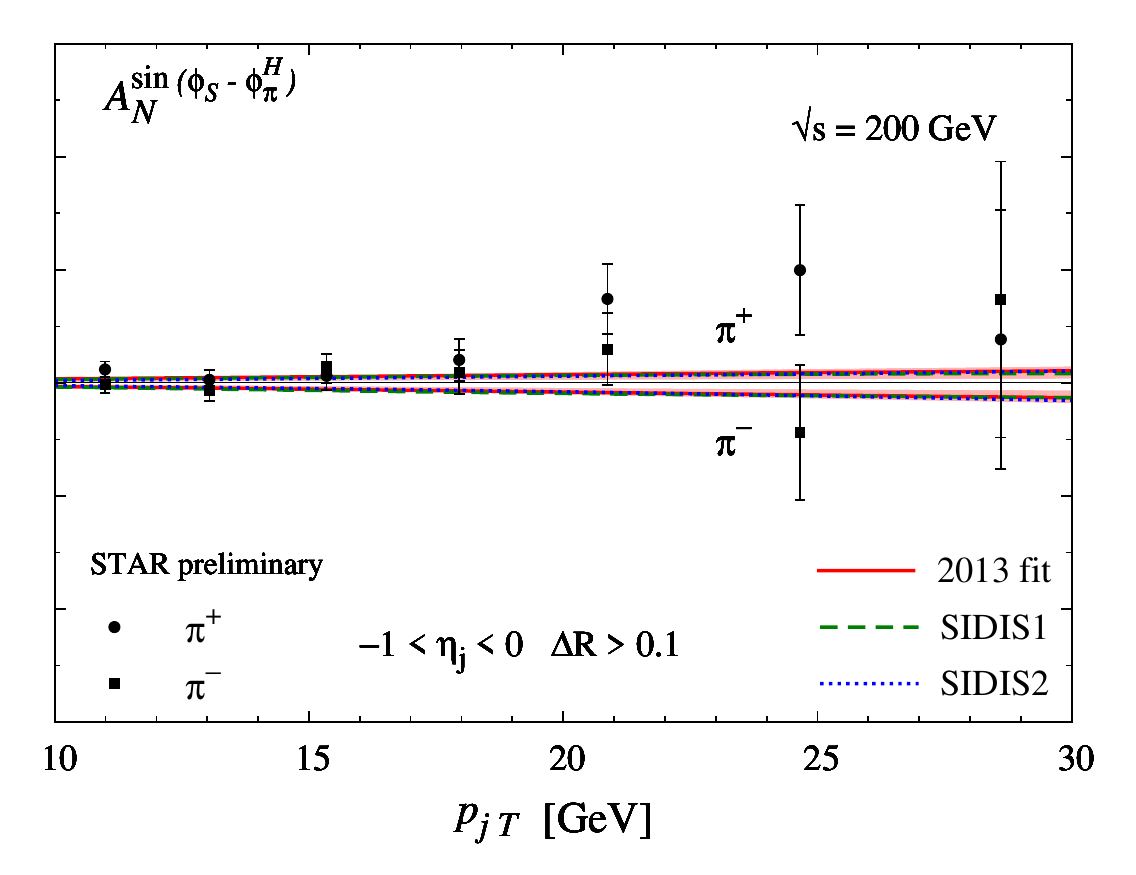}
\caption{The Collins asymmetry  $A_N^{\sin(\phi_{S}-\phi_\pi^H)}$  for the process $p^\uparrow \, p\to {\rm jet}\,
\pi^{\pm} \, X$, as a function of $p_{{\rm j}T}$, at c.m.~energy $\sqrt{s}= 200$ GeV, separately for forward (left panel) and backward (right panel) rapidities, compared with STAR preliminary data~\cite{Adkins:2016uxv}. Estimates are obtained as in the previous figure.}
\label{fig:asy-coll-par-pT}
\end{center}
\end{figure}

In Fig.~\ref{fig:asy-coll-par-pT} we present our predictions for the Collins asymmetry for $\pi^\pm$ production as a function of $p_{{\rm j}T}$ at $\sqrt s=200$~GeV compared with STAR preliminary data. Here we notice a different behaviour, w.r.t.~what discussed above: at large $p_{\rm{j}T}\ge 20$~GeV, in fact, the SIDIS~2 set gives asymmetries larger than those obtained with the SIDIS~1 set and the 2013 fit. The reason is twofold: since the average value of $x_a$ in the polarized proton increases with  $p_{{\rm j}T}$ (for instance at $p_{\rm{j}T}\simeq 20$~GeV $\langle x_a\rangle \simeq 0.3$), from one side the transversity function is probed in the large-$x$ region where it is basically unconstrained. In particular, in the SIDIS~1 set as well as in the 2013 fit it results accidentally more suppressed than for SIDIS~2. From the other side, the role of gluons, both in the distribution and fragmentation functions, becomes less relevant, reducing the differences between the FF sets.

\begin{figure}[ht]
\begin{center}
\includegraphics[width=8.15truecm,angle=0]{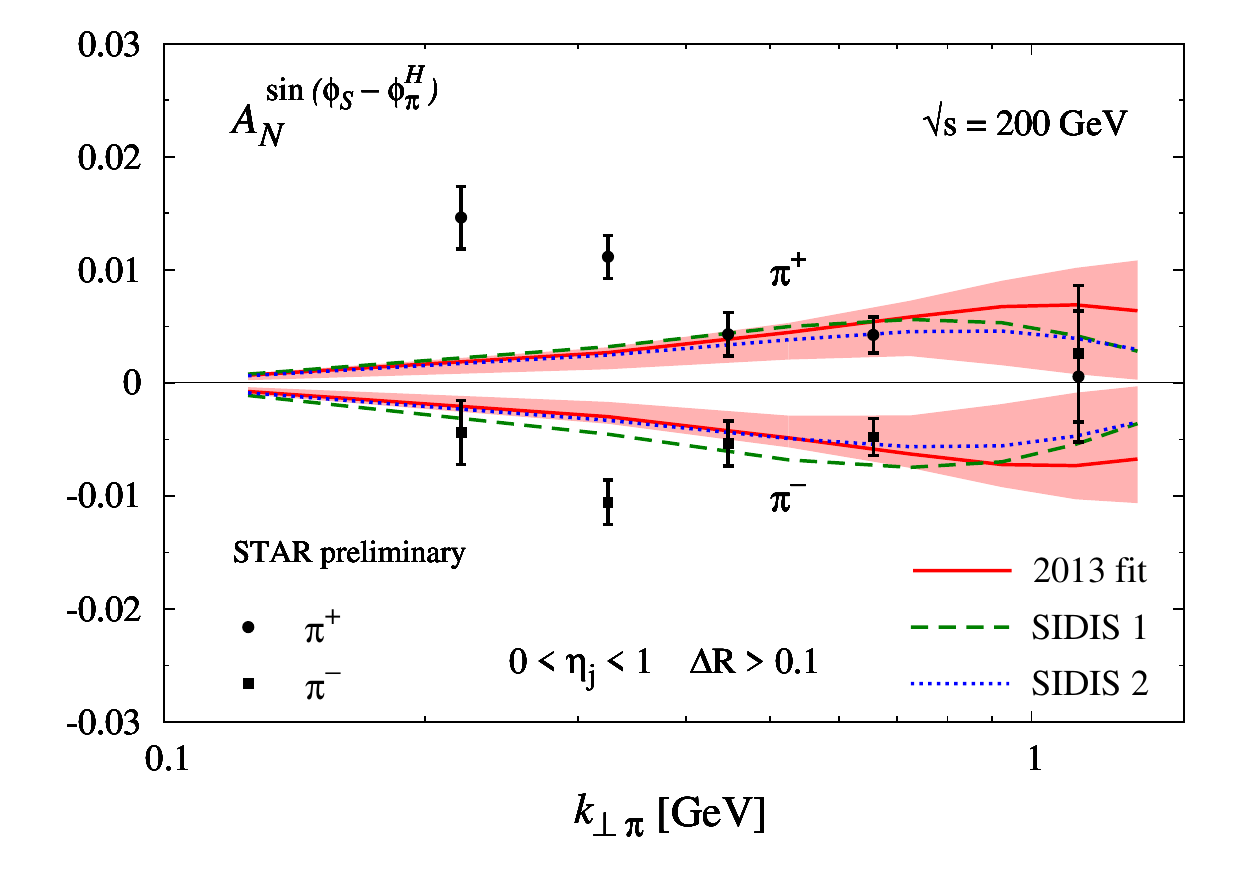}
\includegraphics[width=8.15truecm,angle=0]{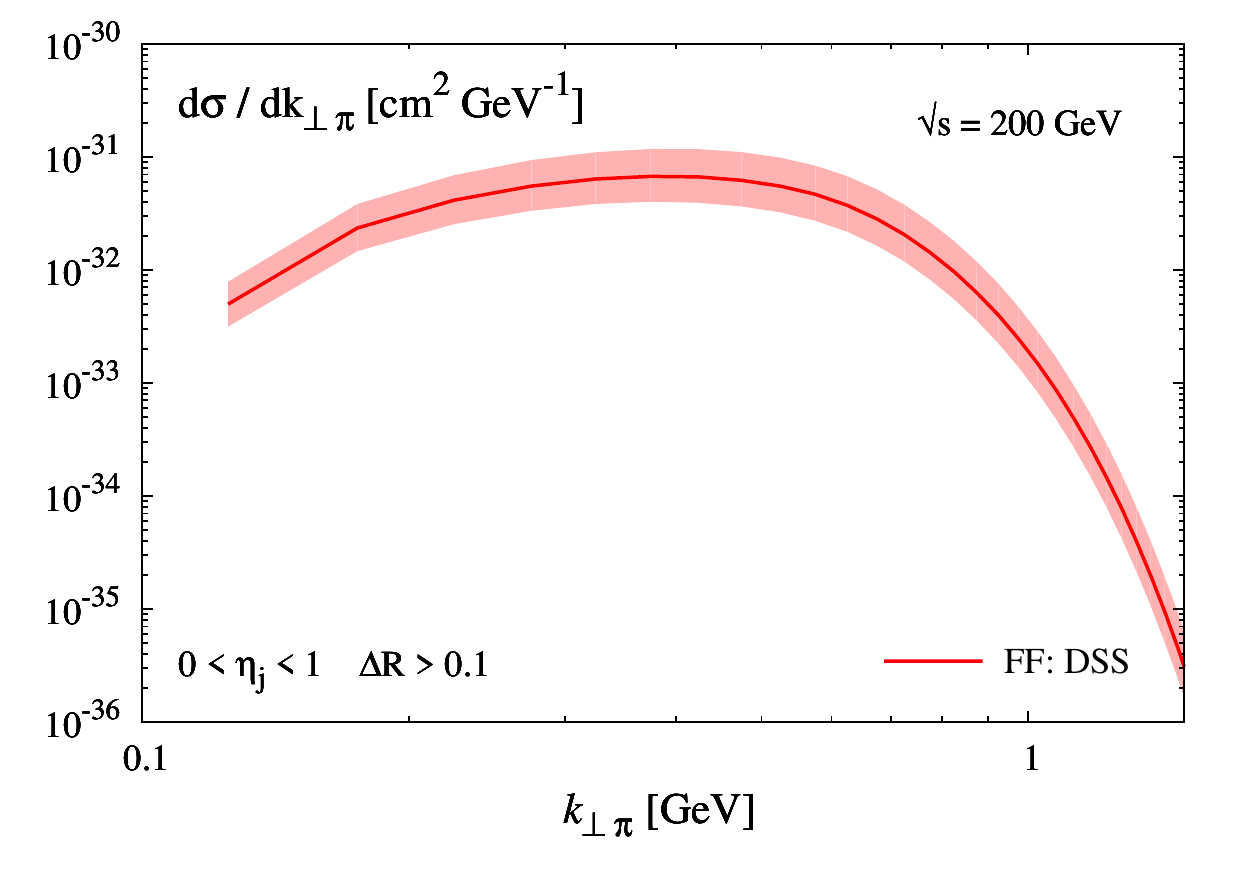}
\caption{The Collins asymmetry $A_N^{\sin(\phi_{S}-\phi_\pi^H)}$ for the process $p^\uparrow \, p\to {\rm jet}\, \pi^{\pm} \, X$ (left panel), and the unpolarised cross section for $\pi^+$ production (right panel), as a function of $k_{\perp \pi}$ (the intrinsic transverse momentum in the fragmentation process, also denoted as $j_T$), at c.m.~energy $\sqrt{s}= 200$ GeV. Preliminary data are from the STAR Collaboration~\cite{Adkins:2016uxv}.
Estimates for the asymmetry are obtained as in the previous figures, while the DSS FF set is used for the calculation of the unpolarised cross section. In such a case we show the uncertainty band obtained varying the factorisation scale between $p_{\rm{j}T}/2$ and $2p_{\rm{j}T}$}.
\label{fig:asy-coll-par-k3}
\end{center}
\end{figure}

In Fig.~\ref{fig:asy-coll-par-k3}, left panel, we show our estimates for the Collins asymmetry for $\pi^\pm$ production as a function of the intrinsic transverse momentum of the pion w.r.t.~the jet direction compared with the STAR preliminary data, still adopting the TMD sets discussed above. For completeness (see also the discussion below), as an example, in the right panel we show the corresponding unpolarised cross section for $\pi^+$ production using the DSS FF set.

We notice that at relatively large $k_{\perp \pi}$  our estimates for the single spin asymmetry are in fair agreement with the data, while at lower values some discrepancies appear.
More precisely, taking into account the status of the data, still preliminary, and the fact that one would expect a better agreement in the low $k_{\perp \pi}$ region, among the possible sources of these discrepancies we could mention: $i)$ a narrower $k_{\perp \pi}$ dependence of the Collins functions (w.r.t.~the ones adopted here), with or without a narrower Gaussian width for the unpolarised TMD FFs; $ii)$ a different $k_{\perp \pi}$ behaviour for favoured and disfavoured TMD FFs; $iii)$ an explicit $z$ dependence in the widths. These changes indeed could shift the maximum of the asymmetry to smaller $k_{\perp \pi}$ values, differentiating also its behaviour for $\pi^+$ and $\pi^-$ production, with a slight impact on the $k_{\perp\pi}$-integrated predictions shown in the previous figures. On the other hand, the same changes should also be checked against the SIDIS and $e^+e^-$ data, from which the Collins FFs have been extracted. This is certainly an important study that, implying the use of an increasing number of free parameters, could be performed only in future extractions of TMDs. For the time being, it is worth reminding that we still do not have at our disposal precise data on the $k_\perp$ dependencies of the unpolarised cross sections and the azimuthal asymmetries in $e^+e^-$ processes, and therefore an accurate knowledge of the explicit transverse momentum dependence of the unpolarised and Collins TMD FFs is still missing. For this reason data on unpolarised cross sections for the process under study here and their comparison with the estimates shown in the right panel of Fig.~\ref{fig:asy-coll-par-k3} could be very useful.

One has also to recall that available information on this dependence is based on the analysis of SIDIS processes (namely multiplicities and azimuthal asymmetries), where it appears to be strongly correlated with the corresponding transverse momentum dependence in the TMD parton distribution functions (see, for instance, Ref.~\cite{Bacchetta:2017gcc}). A preliminary study of the impact of the Gaussian widths in the analysis of SSAs is in progress~\cite{Anselmino:2017xxx} and, due to its relevance in the context of the universality issue, more efforts, both from the experimental and the phenomenological point of view, are definitely necessary. For these reasons the results shown in Fig.~\ref{fig:asy-coll-par-k3} deserve further attention.

Some more general comments are in order:
\begin{itemize}
  \item In our calculations the average intrinsic transverse momentum in the FF, $\langle k_{\perp \pi}\rangle$, is in the range 0.4-0.5 GeV at 200~GeV and 0.3-0.8~GeV at 500~GeV; moreover, the average transverse momentum of the jet, $\langle p_{\rm{j}T} \rangle$, is around  12~GeV at 200~GeV and 25-27~GeV at 500~GeV. This means we are in the proper regime to apply the TMD approach and that one should be sensible to scale evolution effects. Notice that the angular cuts, characterised by the minimum distance of the charged pion from the jet thrust axis, have been chosen to sample the same $x_T$ values ($x_T=2p_{{\rm j}T}/\sqrt s$).
  \item From this analysis we can conclude that the Collins function manifests its universality not only in SIDIS and $e^+e^-$ processes, but also in the azimuthal distribution of pions within jets in hadronic collisions. No compelling factorisation breaking effects emerge.
  \item The fact that the ordinary collinear evolution with the hard scale, as adopted in the extraction of the Collins functions and in the current predictions, is able to describe fairly well the experimental data at very different energy scales suggests a very small role of the more complex TMD evolution. This could be related to a partial cancellation of its effects in a SSA, that is a ratio of cross sections.
  \item The apparent tension between the poor description of the very low $k_{\perp\pi}$ behaviour of the asymmetry for $\pi^+$ production (Fig.~\ref{fig:asy-coll-par-k3}) and the overall good agreement with the data as a function of $z$ and $p_{\rm{j}T}$ can be ascribed to the fact that the $k_{\perp\pi}$-integrated observables are less sensitive to the transverse momentum dependence of the Collins functions.
  \item Concerning the use of the latest extraction~\cite{Anselmino:2015sxa} of the transversity distribution and the Collins FF, within the same scheme (that is including only the DGLAP scale evolution), this would give smaller results for the Collins asymmetry. This is mainly due to the lower value of the Gaussian width for the TMD-FFs adopted in that analysis. This is a delicate issue and a detailed study of the impact of the Gaussian widths in the phenomenological analysis of the Sivers and Collins effects is under way~\cite{Anselmino:2017xxx}.
  \item Another set of data~\cite{Adkins:2016uxv} (lower plot of their Fig.~4), exploring a much larger pion intrinsic transverse momentum w.r.t.~the jet axis, is almost compatible with zero and consistent with our estimates (not shown here).
\end{itemize}

\begin{figure}[ht]
\begin{center}
\hspace*{-8cm}
\includegraphics[width=8.1truecm,angle=0]{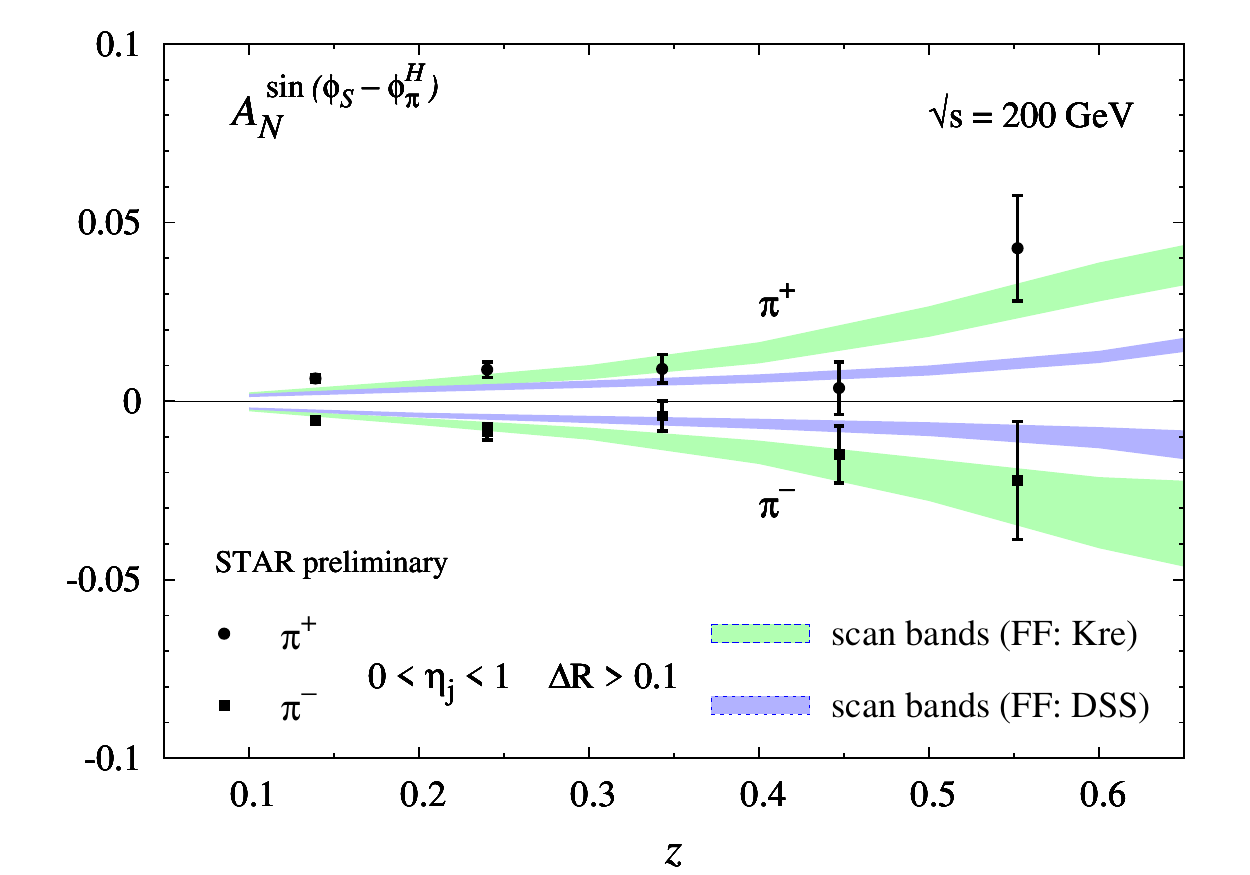}\\
\vspace*{-5.72cm}\hspace*{7cm}
\includegraphics[width=8.1truecm,angle=0]{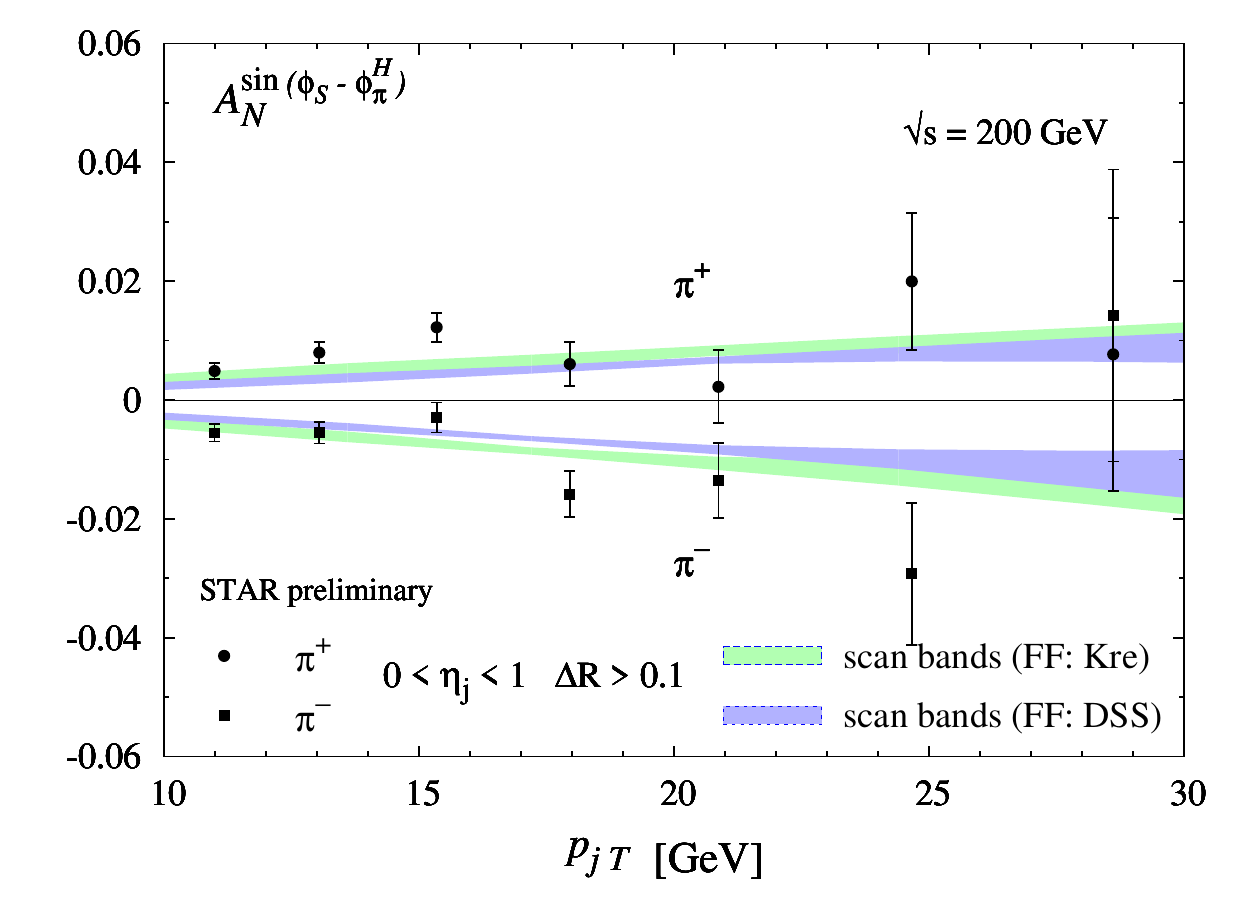}
\caption{The Collins asymmetry  $A_N^{\sin(\phi_{S}-\phi_\pi^H)}$  for the process $p^\uparrow \, p\to {\rm jet}\,
\pi^{\pm} \, X$, at c.m.~energy $\sqrt{s}= 200$ GeV, as a function of $z$ (left panel) and $p_{\rm{j}T}$, (right panel), compared with STAR preliminary data~\cite{Adkins:2016uxv}. The shaded areas represent the envelope of all possible values of the asymmetry obtained following the procedure of Ref.~\cite{Anselmino:2012rq}, for two different FF sets.}
\label{fig:asy-coll-ppscan}
\end{center}
\end{figure}

Before concluding this section we would like to comment further on the impact of the lack of knowledge of the transversity distribution at large $x$. This, for instance, could be relevant in order to properly assess potential factorisation breaking and/or TMD evolution effects. To this end in Fig.~\ref{fig:asy-coll-ppscan} we show some estimates obtained by sampling the parameter controlling the large-$x$ behaviour of the transversity distribution, as described in Ref.~\cite{Anselmino:2012rq}. Notice that in such a case the shaded areas (referred to as \emph{scan bands} in Ref.~\cite{Anselmino:2012rq}) represent only the envelope of all possible values of the asymmetry obtained following the above procedure. They are smaller than the corresponding statistical uncertainty bands shown in the left panels of Figs.~\ref{fig:asy-coll-par-z} and \ref{fig:asy-coll-par-pT}, showing that such bands are not only due to the uncertainty on the transversity distribution at large $x$. Without entering into further details, and taking into account the above considerations, these results allow us to confirm once again our conclusions on the universality of the Collins function and on the almost negligible role of TMD evolution effects.

\subsection{Collins-like azimuthal asymmetry}
\label{collins-like}

\begin{figure}[t!]
\begin{center}
\includegraphics[width=8.25truecm,angle=0]{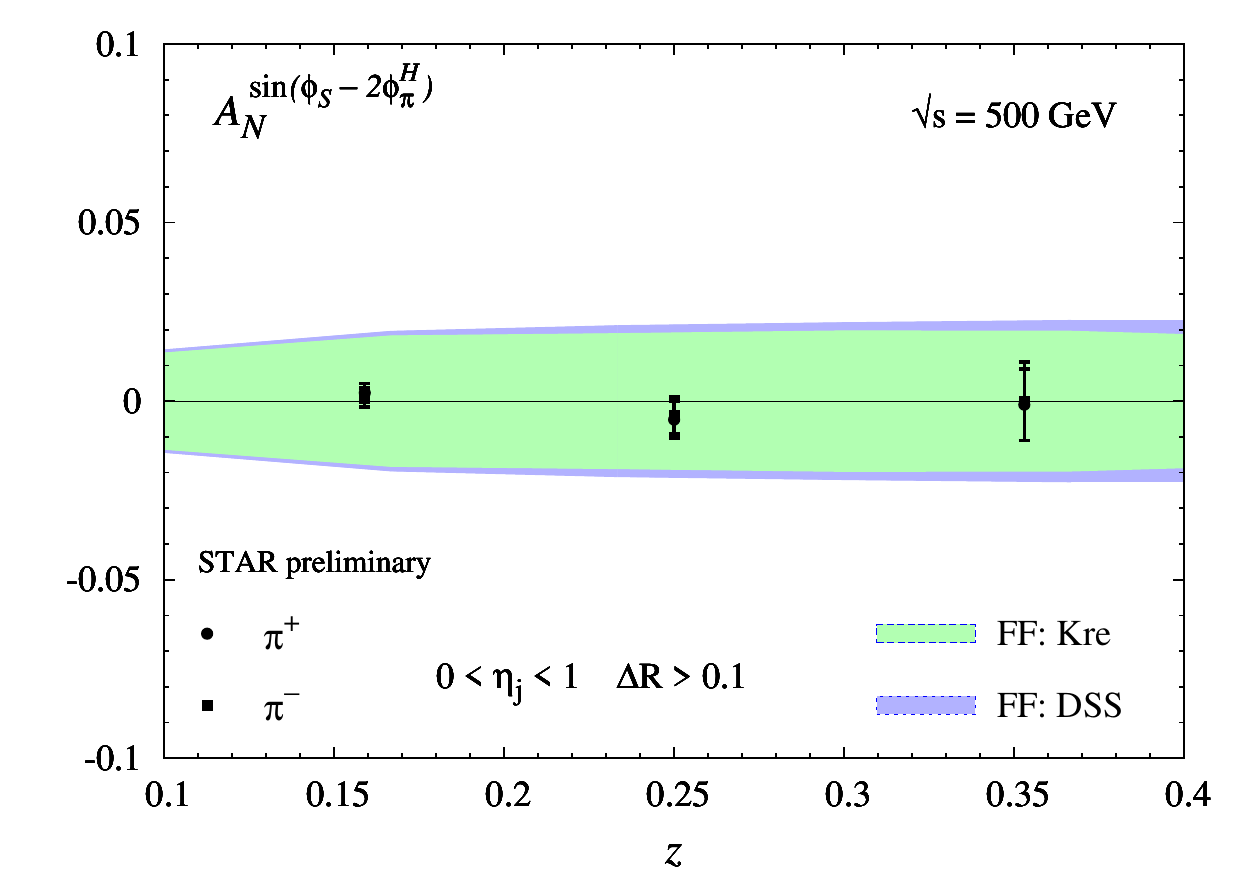}\hspace*{-0.5 cm}
\includegraphics[width=7.5truecm,angle=0]{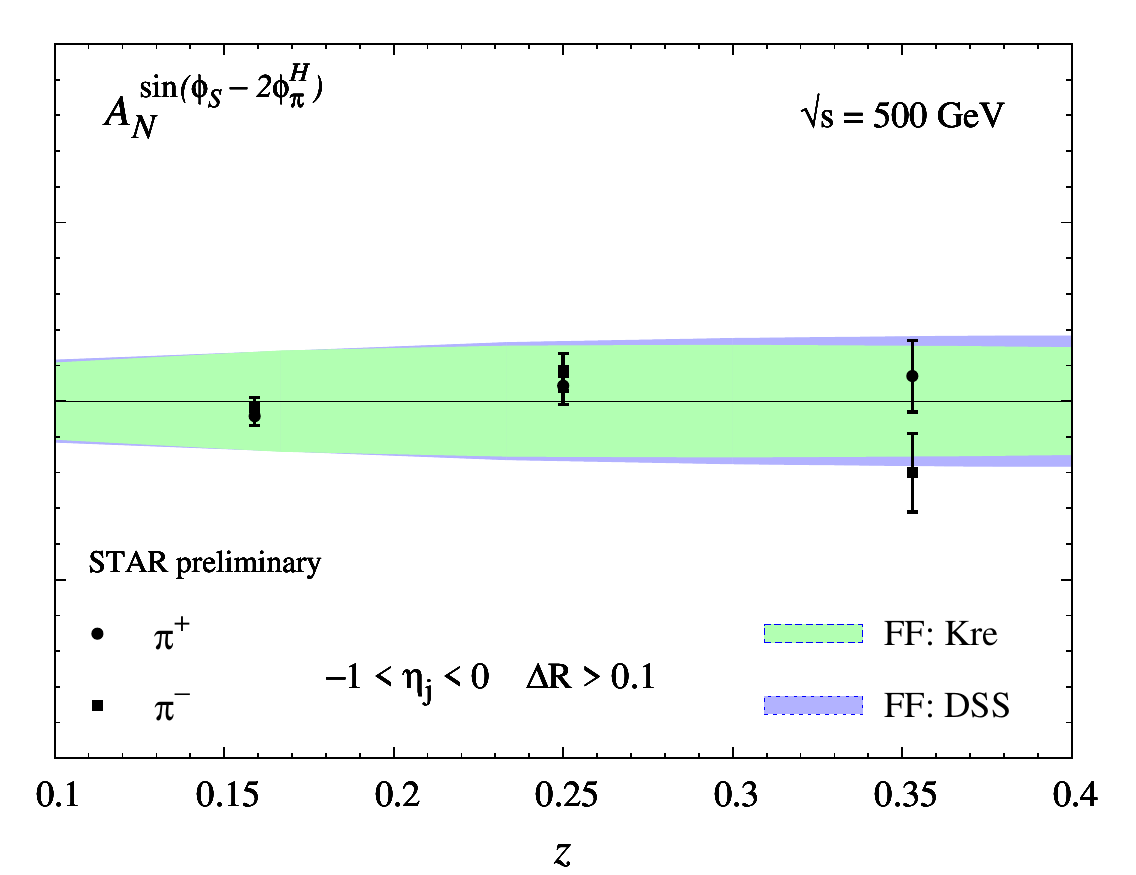}
\caption{The Collins-like asymmetry  $A_N^{\sin(\phi_{S}-2\phi_\pi^H)}$  for the process $p^\uparrow \, p\to {\rm jet}\,
\pi^{\pm} \, X$, as a function of $z$, at c.m.~energy
$\sqrt{s}= 200$ GeV, separately for forward (left panel) and backward (right panel) rapidities, compared with STAR preliminary data~\cite{Drachenberg:2014txa,Drachenberg:2014jla}. Estimates are computed by integrating over the other variables and imposing the experimental cuts as summarized in the legend (see also Table 1). Bands represent the maximum values in size of the SSA obtained by saturating the positivity bounds of the two gluon TMDs and adopting two different FF sets.}
\label{fig:asy-coll-like}
\end{center}
\end{figure}

Following the general results presented in Ref.~\cite{D'Alesio:2010am}, that show how to correlate different angular modulations to different TMDs, STAR has extracted several other angular modulations~\cite{Drachenberg:2014txa,Drachenberg:2014jla}. One example is the Collins-like asymmetry $A_{UT}^{\sin(\phi_S-2\phi_\pi^H)}$ as a function of $z$, integrated over the full acceptance and separated in forward and backward scattering relative to the polarised beam. In Fig.~\ref{fig:asy-coll-like} we present these STAR preliminary data, almost compatible with zero, together with our maximised estimates obtained by saturating the positivity bounds\footnote{We maximise the first $k_\perp$-moment of the transverse momentum dependent factor and use for the $x$ and $z$ dependent parts the corresponding unpolarised gluon PDF and FF.} of the two unknown TMDs related to the linear polarisation of gluons (inside the polarised proton and in the fragmentation process). It is worth noticing that for this data set the average value of the momentum fraction inside the protons is around 0.05 and $\langle p_{\rm{j}T}\rangle = $ 7-8 GeV.

As one can see, the maximised prediction suggests a possible upper limit of $\sim$ 2\% (independently of the FF set), while the present data fall well below this maximum, with the best precision at lower values of $z$. Thus, these data represent the first phenomenological constraint on model predictions utilizing linearly polarised gluons, beyond their positivity bounds. We can safely say that these experimental results imply that the {\em product} of these two unknown linearly polarised gluon TMDs cannot be larger than 20-25\% the product of their positivity bounds (focusing on the lowest $z$ bin).

\section{Conclusions}
\label{conclusions}
In this letter we have studied the Collins and Collins-like azimuthal asymmetries in the distribution of hadrons within jets produced in polarised proton-proton collisions. By applying a TMD approach at leading order, with inclusion of spin and transverse momentum effects, we have shown how the preliminary data of the Collins asymmetry collected by the STAR Collaboration at RHIC could be described in terms of a universal Collins function, with only some discrepancies in the description of their $k_{\perp\pi}$ behaviour. If confirmed by data, this would require a better understanding of the $k_\perp$ dependencies of the unpolarised and the Collins TMD FFs as extracted from current phenomenological analyses.
These results, obtained adopting available estimates of the transversity distribution and the Collins FF, as extracted from independent fits of the azimuthal asymmetries observed in SIDIS and $e^+e^-$ processes, with standard DGLAP QCD evolution, give indeed an overall good description of the STAR data in different kinematical configurations. No indication of universality-breaking effects emerges and, quite interestingly, no compelling TMD evolution effects appear. With some caution, this could be considered as the first phenomenological evidence for the universality of the Collins function in hadron-hadron collisions.

The Collins-like asymmetry, involving linearly polarised gluons, both inside the polarised proton and in the fragmentation process, has also been considered. In such a case the STAR preliminary data allow for the first phenomenological constraint on the size of these TMDs.

Further study is certainly necessary, but these results open a new window in the field of TMD effects in hadronic collisions.

Upon completion of this analysis we became aware of a similar study, including also TMD evolution effects, performed in Ref.~\cite{Kang:2017xxx}. Their results, in agreement with our findings, support our conclusions.

\section*{Acknowledgments}
We would like to thank Kevin Adkins, Jim Drachenberg, Renee Fatemi and  Carl Gagliardi for useful information on the STAR preliminary data. C.P.~is supported by the European Research Council (ERC) under the European Union's Horizon 2020 research and innovation programme (grant agreement No.~647981, 3DSPIN).

\section*{References}


\begin{thebibliography}{10}
\expandafter\ifx\csname url\endcsname\relax
  \def\url#1{\texttt{#1}}\fi
\expandafter\ifx\csname urlprefix\endcsname\relax\def\urlprefix{URL }\fi
\expandafter\ifx\csname href\endcsname\relax
  \def\href#1#2{#2} \def\path#1{#1}\fi

\bibitem{D'Alesio:2007jt}
U.~D'Alesio, F.~Murgia, {Azimuthal and single spin asymmetries in hard
  scattering processes}, Prog. Part. Nucl. Phys. 61 (2008) 394--454.
\newblock \href {http://arxiv.org/abs/0712.4328} {\path{arXiv:0712.4328}},
  \href {http://dx.doi.org/10.1016/j.ppnp.2008.01.001}
  {\path{doi:10.1016/j.ppnp.2008.01.001}}.

\bibitem{Barone:2010zz}
V.~Barone, F.~Bradamante, A.~Martin, {Transverse-spin and transverse-momentum
  effects in high-energy processes}, Prog. Part. Nucl. Phys. 65 (2010)
  267--333.
\newblock \href {http://arxiv.org/abs/1011.0909} {\path{arXiv:1011.0909}},
  \href {http://dx.doi.org/10.1016/j.ppnp.2010.07.003}
  {\path{doi:10.1016/j.ppnp.2010.07.003}}.

\bibitem{Aschenauer:2015ndk}
E.~C. Aschenauer, U.~D'Alesio, F.~Murgia, {TMDs and SSAs in hadronic
  interactions}, Eur. Phys. J. A52 (2016) 156.
\newblock \href {http://arxiv.org/abs/1512.05379} {\path{arXiv:1512.05379}},
  \href {http://dx.doi.org/10.1140/epja/i2016-16156-4}
  {\path{doi:10.1140/epja/i2016-16156-4}}.

\bibitem{Collins:2011zzd}
J.~Collins, {Foundations of perturbative QCD}, Cambridge University Press,
  2011.

\bibitem{GarciaEchevarria:2011rb}
M.~G. Echevarria, A.~Idilbi, I.~Scimemi, {Factorization theorem for Drell-Yan
  at low $q_T$ and transverse momentum distributions on-the-light-cone}, JHEP
  07 (2012) 002.
\newblock \href {http://arxiv.org/abs/1111.4996} {\path{arXiv:1111.4996}},
  \href {http://dx.doi.org/10.1007/JHEP07(2012)002}
  {\path{doi:10.1007/JHEP07(2012)002}}.

\bibitem{Collins:2017oxh}
J.~Collins, T.~C. Rogers, {Connecting different TMD factorization formalisms in
  QCD}. \href {http://arxiv.org/abs/1705.07167} {\path{arXiv:1705.07167}}.

\bibitem{Kouvaris:2006zy}
C.~Kouvaris, J.-W. Qiu, W.~Vogelsang, F.~Yuan, {Single transverse-spin
  asymmetry in high transverse momentum pion production in $p p$ collisions},
  Phys. Rev. D74 (2006) 114013.
\newblock \href {http://arxiv.org/abs/hep-ph/0609238}
  {\path{arXiv:hep-ph/0609238}}, \href
  {http://dx.doi.org/10.1103/PhysRevD.74.114013}
  {\path{doi:10.1103/PhysRevD.74.114013}}.

\bibitem{Kanazawa:2015ajw}
K.~Kanazawa, Y.~Koike, A.~Metz, D.~Pitonyak, M.~Schlegel, {Operator constraints
  for twist-3 functions and Lorentz invariance properties of twist-3
  observables}, Phys. Rev. D93 (2016) 054024.
\newblock \href {http://arxiv.org/abs/1512.07233} {\path{arXiv:1512.07233}},
  \href {http://dx.doi.org/10.1103/PhysRevD.93.054024}
  {\path{doi:10.1103/PhysRevD.93.054024}}.

\bibitem{Sivers:1989cc}
D.~W. Sivers, {Single spin production asymmetries from the hard scattering of
  point-like constituents}, Phys. Rev. D41 (1990) 83.
\newblock \href {http://dx.doi.org/10.1103/PhysRevD.41.83}
  {\path{doi:10.1103/PhysRevD.41.83}}.

\bibitem{Collins:1992kk}
J.~C. Collins, {Fragmentation of transversely polarized quarks probed in
  transverse momentum distributions}, Nucl. Phys. B396 (1993) 161--182.
\newblock \href {http://arxiv.org/abs/hep-ph/9208213}
  {\path{arXiv:hep-ph/9208213}}, \href
  {http://dx.doi.org/10.1016/0550-3213(93)90262-N}
  {\path{doi:10.1016/0550-3213(93)90262-N}}.

\bibitem{Collins:2004nx}
J.~C. Collins, A.~Metz, {Universality of soft and collinear factors in
  hard-scattering factorization}, Phys. Rev. Lett. 93 (2004) 252001.
\newblock \href {http://arxiv.org/abs/hep-ph/0408249}
  {\path{arXiv:hep-ph/0408249}}, \href
  {http://dx.doi.org/10.1103/PhysRevLett.93.252001}
  {\path{doi:10.1103/PhysRevLett.93.252001}}.

\bibitem{Anselmino:2012rq}
M.~Anselmino, M.~Boglione, U.~D'Alesio, E.~Leader, S.~Melis, F.~Murgia,
  A.~Prokudin, {Role of Collins effect in the single spin asymmetry $A_N$ in
  $p^\uparrow p \to h X$ processes}, Phys. Rev. D86 (2012) 074032.
\newblock \href {http://arxiv.org/abs/1207.6529} {\path{arXiv:1207.6529}},
  \href {http://dx.doi.org/10.1103/PhysRevD.86.074032}
  {\path{doi:10.1103/PhysRevD.86.074032}}.

\bibitem{Anselmino:2013rya}
M.~Anselmino, M.~Boglione, U.~D'Alesio, S.~Melis, F.~Murgia, A.~Prokudin,
  {Sivers effect and the single spin asymmetry $A_N$ in $p^\uparrow p \to h X$
  processes}, Phys. Rev. D88 (2013) 054023.
\newblock \href {http://arxiv.org/abs/1304.7691} {\path{arXiv:1304.7691}},
  \href {http://dx.doi.org/10.1103/PhysRevD.88.054023}
  {\path{doi:10.1103/PhysRevD.88.054023}}.

\bibitem{Rogers:2013zha}
T.~C. Rogers, {Extra spin asymmetries from the breakdown of
  transverse-momentum-dependent factorization in hadron-hadron collisions},
  Phys. Rev. D88 (2013) 014002.
\newblock \href {http://arxiv.org/abs/1304.4251} {\path{arXiv:1304.4251}},
  \href {http://dx.doi.org/10.1103/PhysRevD.88.014002}
  {\path{doi:10.1103/PhysRevD.88.014002}}.

\bibitem{Collins:2007nk}
  J.~Collins and J.~W.~Qiu,
  {$k_{T}$ factorization is violated in production of high-transverse-momentum particles in hadron-hadron collisions},
  Phys. Rev. D75 (2007) 114014.
  \newblock \href {http://arxiv.org/abs/0705.2141} {\path{arXiv:0705.2141}},
  \href {http://dx.doi.org/10.1103/PhysRevD.75.114014}
  {\path{doi:10.1103/PhysRevD.75.114014}}.


\bibitem{D'Alesio:2011mc}
U.~D'Alesio, L.~Gamberg, Z.-B. Kang, F.~Murgia, C.~Pisano, {Testing the process
  dependence of the {S}ivers function via hadron distributions inside a jet},
  Phys. Lett. B704 (2011) 637--640.
\newblock \href {http://arxiv.org/abs/1108.0827} {\path{arXiv:1108.0827}},
  \href {http://dx.doi.org/10.1016/j.physletb.2011.09.067}
  {\path{doi:10.1016/j.physletb.2011.09.067}}.

\bibitem{Yuan:2007nd}
F.~Yuan, {Azimuthal asymmetric distribution of hadrons inside a jet at hadron
  collider}, Phys. Rev. Lett. 100 (2008) 032003.
\newblock \href {http://arxiv.org/abs/0709.3272} {\path{arXiv:0709.3272}},
  \href {http://dx.doi.org/10.1103/PhysRevLett.100.032003}
  {\path{doi:10.1103/PhysRevLett.100.032003}}.

\bibitem{D'Alesio:2010am}
U.~D'Alesio, F.~Murgia, C.~Pisano, {Azimuthal asymmetries for hadron
  distributions inside a jet in hadronic collisions}, Phys. Rev. D83 (2011)
  034021.
\newblock \href {http://arxiv.org/abs/1011.2692} {\path{arXiv:1011.2692}},
  \href {http://dx.doi.org/10.1103/PhysRevD.83.034021}
  {\path{doi:10.1103/PhysRevD.83.034021}}.

\bibitem{DAlesio:2013cfy}
U.~D'Alesio, F.~Murgia, C.~Pisano, {Collins and Sivers effects in $p^\uparrow p
  \to$ jet $\pi X$: Universality and process dependence}, Phys. Part. Nucl.
  45 (2014) 676--691.
\newblock \href {http://arxiv.org/abs/1307.4880} {\path{arXiv:1307.4880}},
  \href {http://dx.doi.org/10.1134/S1063779614040054}
  {\path{doi:10.1134/S1063779614040054}}.


\bibitem{Adkins:2016uxv}
J.~K. Adkins, J.~L. Drachenberg, {Azimuthal single-spin asymmetries of charged
  pions in jets in $\sqrt s =$ 200 GeV $p^\uparrow p$ collisions at STAR}, Int.
  J. Mod. Phys. Conf. Ser. 40 (2016) 1660040.
\newblock \href {http://dx.doi.org/10.1142/S2010194516600405}
  {\path{doi:10.1142/S2010194516600405}}.

\bibitem{Drachenberg:2014txa}
J.~L. Drachenberg,
  \href{http://inspirehep.net/record/1375601/files/78.pdf}{{Constraining
  Transversity and nucleon transverse-polarization structure through
  polarized-proton collisions at STAR}}, in: {Proceedings, 20th International
  Conference on Particles and Nuclei (PANIC 14): Hamburg, Germany, August
  24-29, 2014}, pp. 181--184.
\newblock \href {http://dx.doi.org/10.3204/DESY-PROC-2014-04/78}
  {\path{doi:10.3204/DESY-PROC-2014-04/78}}.
\newline\urlprefix\url{http://inspirehep.net/record/1375601/files/78.pdf}

\bibitem{Drachenberg:2014jla}
J.~L. Drachenberg, {Transverse single-spin asymmetries from $p^{\uparrow} + p
  \to$ jet + X and $p^{\uparrow} + p \to$ jet $+ \pi^{\pm} + X$ at $\sqrt{s}$ =
  500 GeV at RHIC}, EPJ Web Conf. 73 (2014) 02009.
\newblock \href {http://dx.doi.org/10.1051/epjconf/20147302009}
  {\path{doi:10.1051/epjconf/20147302009}}.

\bibitem{Boer:2016fqd}
D.~Boer, P.~J. Mulders, C.~Pisano, J.~Zhou, {Asymmetries in heavy quark pair
  and dijet production at an EIC}, JHEP 08 (2016) 001.
\newblock \href {http://arxiv.org/abs/1605.07934} {\path{arXiv:1605.07934}},
  \href {http://dx.doi.org/10.1007/JHEP08(2016)001}
  {\path{doi:10.1007/JHEP08(2016)001}}.



\bibitem{Anselmino:2007fs}
M.~Anselmino, M.~Boglione, U.~D'Alesio, A.~Kotzinian, F.~Murgia, A.~Prokudin,
  C.~T\"urk, {Transversity and Collins functions from SIDIS and $e^+ e^-$ data},
  Phys. Rev. D75 (2007) 054032.
\newblock \href {http://arxiv.org/abs/hep-ph/0701006}
  {\path{arXiv:hep-ph/0701006}}, \href
  {http://dx.doi.org/10.1103/PhysRevD.75.054032}
  {\path{doi:10.1103/PhysRevD.75.054032}}.

\bibitem{Anselmino:2008jk}
M.~Anselmino, M.~Boglione, U.~D'Alesio, A.~Kotzinian, F.~Murgia, A.~Prokudin,
  S.~Melis, {Update on transversity and Collins functions from SIDIS and $e^+ e^-$
  data}, Nucl. Phys. Proc. Suppl. 191 (2009) 98--107.
\newblock \href {http://arxiv.org/abs/0812.4366} {\path{arXiv:0812.4366}},
  \href {http://dx.doi.org/10.1016/j.nuclphysbps.2009.03.117}
  {\path{doi:10.1016/j.nuclphysbps.2009.03.117}}.

\bibitem{anselmino:2013vqa}
M.~Anselmino, M.~Boglione, U.~D'Alesio, S.~Melis, F.~Murgia, A.~Prokudin,
  {Simultaneous extraction of transversity and Collins functions from new SIDIS
  and $e^+e^-$ data}, Phys. Rev. D87 (2013) 094019.
\newblock \href {http://arxiv.org/abs/1303.3822} {\path{arXiv:1303.3822}},
  \href {http://dx.doi.org/10.1103/PhysRevD.87.094019}
  {\path{doi:10.1103/PhysRevD.87.094019}}.


\bibitem{Anselmino:2008sga}
M.~Anselmino, M.~Boglione, U.~D'Alesio, A.~Kotzinian, S.~Melis, F.~Murgia,
  A.~Prokudin, {Sivers effect for pion and kaon production in semi-inclusive
  deep inelastic scattering}, Eur. Phys. J. A39 (2009) 89--100.
\newblock \href {http://arxiv.org/abs/0805.2677} {\path{arXiv:0805.2677}},
  \href {http://dx.doi.org/10.1140/epja/i2008-10697-y}
  {\path{doi:10.1140/epja/i2008-10697-y}}.

\bibitem{Anselmino:2015sxa}
M.~Anselmino, M.~Boglione, U.~D'Alesio, J.~O. Gonzalez~Hernandez, S.~Melis,
  F.~Murgia, A.~Prokudin, {Collins functions for pions from SIDIS and new
  $e^+e^-$ data: a first glance at their transverse momentum dependence}, Phys.
  Rev. D92 (2015) 114023.
\newblock \href {http://arxiv.org/abs/1510.05389} {\path{arXiv:1510.05389}},
  \href {http://dx.doi.org/10.1103/PhysRevD.92.114023}
  {\path{doi:10.1103/PhysRevD.92.114023}}.


\bibitem{Kretzer:2000yf}
S.~Kretzer, {Fragmentation functions from flavor inclusive and flavor tagged $e^+
  e^-$ annihilations}, Phys. Rev. D62 (2000) 054001.
\newblock \href {http://arxiv.org/abs/hep-ph/0003177}
  {\path{arXiv:hep-ph/0003177}}, \href
  {http://dx.doi.org/10.1103/PhysRevD.62.054001}
  {\path{doi:10.1103/PhysRevD.62.054001}}.

\bibitem{deFlorian:2007aj}
D.~de~Florian, R.~Sassot, M.~Stratmann, {Global analysis of fragmentation
  functions for pions and kaons and their uncertainties}, Phys. Rev. D75 (2007)
  114010.
\newblock \href {http://arxiv.org/abs/hep-ph/0703242}
  {\path{arXiv:hep-ph/0703242}}, \href
  {http://dx.doi.org/10.1103/PhysRevD.75.114010}
  {\path{doi:10.1103/PhysRevD.75.114010}}.

\bibitem{Kaufmann:2015hma}
  T.~Kaufmann, A.~Mukherjee and W.~Vogelsang,
  {Hadron fragmentation inside jets in hadronic collisions}
  Phys. Rev. D92 (2015)  054015.
\newblock \href {http://arxiv.org/abs/1506.01415} {\path{arXiv:1506.01415}},
  \href {http://dx.doi.org/10.1103/PhysRevD.92.054015}
  {\path{doi:10.1103/PhysRevD.92.054015}}.


\bibitem{Kang:2017glf}
Z.-B. Kang, X.~Liu, F.~Ringer, H.~Xing, {The transverse momentum distribution
  of hadrons within jets}. \href {http://arxiv.org/abs/1705.08443}
  {\path{arXiv:1705.08443}}.

\bibitem{Bacchetta:2017gcc}
  A.~Bacchetta, F.~Delcarro, C.~Pisano, M.~Radici and A.~Signori,
  {Extraction of partonic transverse momentum distributions from semi-inclusive deep-inelastic scattering, Drell-Yan and Z-boson production}  JHEP {\bf 1706} (2017) 081.
\newblock  \href {http://arxiv.org/abs/1703.10157} {\path{arXiv:1703.10157}},
 \href {http://dx.doi.org/10.1007/JHEP06(2017)081}
 {\path{doi:10.1007/JHEP06(2017)081}}.

\bibitem{Anselmino:2017xxx}
M.~Anselmino, M.~Boglione, U.~D'Alesio, F.~Murgia, A.~Prokudin, in preparation.

\bibitem{Kang:2017xxx}
Z.-B. Kang, A.~Prokudin, F.~Ringer, F.~Yuan, {Collins azimuthal asymmetries of
  hadron production inside jets}. \href {http://arxiv.org/abs/1707.00913}
  {\path{arXiv:1707.00913}}.

\end{thebibliography}

\end{document}